\documentclass[preprint,showpacs,preprintnumbers,prb]{revtex4} 
\usepackage{graphicx}
\usepackage{amsmath}
\usepackage{times}
\usepackage{epsfig}
\usepackage{color}
\usepackage{ulem}   
\begin{document}
\def\salto{\vskip 1cm} \def\lag{\langle} \def\rag{\rangle}
\newcommand{\redit}[1]{\textcolor{red}{#1}}
\newcommand{\blueit}[1]{\textcolor{blue}{#1}}
\newcommand{\magit}[1]{\textcolor{magenta}{#1}}
\newcommand{\mylabel  }[1]{\label{#1} 
}
\title{Superconducting instability of a non-magnetic metallic band in an antiferromagnetic background}
 
\author{Fernando Agust\'in Reboredo}    

\affiliation {Materials Science and Technology Division, Oak Ridge
National Laboratory, Oak Ridge, TN 37831, USA}
  
\begin{abstract}
It is shown that a non-magnetic metallic band in the presence of an antiferromagnetic background coupled only by the exchange 
interaction develops a superconducting
instability similar to the one described by BCS theory plus additional terms that strongly renormalize the spin excitation spectra. 
A Bardeen-Pines-like hole-hole interaction Hamiltonian, mediated by magnetic excitations, is deduced from
a microscopic model of a fermion band and a spin band that interact with each other only via the exchange interaction.
The theory shows  the appearance of an attractive interaction when the Fermi velocity in 
the non-magnetic band is larger than
the magnon velocity in the magnetic band. The electron-magnon scattering is suppressed by the appearance of a spin gap simultaneously with the superconducting state. 
Although this model may well describe a general class of materials  to be discovered,   the possibility that this theory could describe superconducting cuprates 
is discussed. 
\end{abstract}
\date{\today}
\maketitle

\section{Introduction}
The textbooks classify  materials according to their magnetic behavior as: diamagnetic, paramagnetic, ferromagnetic, ferrimagnetic 
and antiferromagnetic.   Materials can also be insulators or metals. Magnetic materials that are metallic are described as itinerant magnets or antiferromagnets. 
This work discusses the properties of  a general class of materials where an antiferromagnetic insulating band and a non-magnetic metallic band coexist 
and  interact mainly via exchange coupling.  

The electron-phonon interaction is well known to induce a superconducting state in some metals.\cite{BCS,schrieffer83} This type 
of superconductivity is often denoted in the literature as ``conventional superconductivity".  
There has been an intense debate in the literature on the  ``unconventional superconductivity" observed 
in layered structures such as in the cuprates~\cite{bednorz86} or pnicitides.~\cite{kamilhara06} 
  There is significant experimental evidence that suggests that antiferromagnetism and superconductivity are related. In particular,  
  the phase diagram as function of doping shows that superconductivity emerges as antiferromagnetism is suppressed.  Therefore, it is 
  often argued that an electronic mechanism makes the cuprates superconduct and 
   that the same mechanism that suppresses antiferromagnetism gives rise to superconductivity. 
 
 Most theoretical work on the cuprates case is focused on the Cu-O layers. The prevailing view is that the carriers introduced by doping
  go to those planes.  Many have argued in favor of a mechanism 
in which the same electronic bands  that cause antiferromagnetism become superconducting.  Much of the early literature argued that the ``Mott" physics that makes the material an insulator~\cite{imada98,lee06} was also responsible for superconductivity. Since the discovery of superconductivity in materials of the 
pnicitide family,~\cite{kamilhara06} the idea that a Mott insulator is an essential ingredient for unconventional 
superconductivity has lost some ground. Antiferromagnetism, however, remains a common feature of the parent compounds of cuprates and pnictides. 

The cuprates are doped in the so called reservoir regions.~\cite{eisaki04} Doping
directly the CuO$_2$ planes has been shown\cite{pan00} to reduce the critical temperature. The prevailing theoretical point of view is that
no carriers are left behind: all the holes or electrons\cite{tokura89}
travel from the reservoir regions to the CuO$_2$ planes. However, the transfer of charge from the reservoir to the CuO$_2$ planes 
must create a 
repulsive Hartree potential. For typical doping densities (10\%) and typical lattice constants, this repulsive potential can be 
estimated to be of the order of 2-3 eV. Therefore, it is possible that some carriers are left behind in the reservoir. In these conditions, an equilibrium 
between the electronic chemical potential at the bands in the planes and the reservoir must be reached. It is interesting to know if there are conditions where the reservoir can contribute to superconductivity. If more than one band participates in unconventional superconductivity a theory that can account for it must be developed as in the BCS case.~\cite{suhl59}

Since the discovery of unconventional superconductivity,~\cite{bednorz86} early works in the field showed that the spin susceptibility may play an important
 role.~\cite{scalapino86,anderson87,schrieffer88,chen88,schmalian99}  The coupling of the collective spin excitations with the charge degrees of freedom within the same band was shown~\cite{abanov01} to provide a mechanism for superconductivity. A {\it d} wave symmetry for the superconducting ground state can be obtained via coupling with the spin excitations.~\cite{monthoux92} Similar {\it d} wave superconductivity was found in a model system where spins move in a lattice but they cannot occupy the same site (t-J model).\cite{poilblanc94} 
 
 It seems obvious that if the coupling of the spin degrees of freedom of a band with the charge degrees of freedom of the same band provides a mechanism for superconductivity, a similar coupling may also cause superconductivity if different bands are involved. 
 Therefore, a class of new materials that may superconduct must be considered. In this class the charge degrees of freedom of one band are coupled by the exchange interaction with the spin degrees of freedom of a different band.  It is crucial to know
whether superconductivity could occur under conditions different from those which have been explored theoretically and experimentally so far, 
because the discovery of a material that could superconduct at room temperature would certainly revolutionize technology by significantly reducing energy wasted in heat. 

This paper describes and studies a general type of model of a  non-magnetic metallic band, coupled by the exchange interaction to an antiferromagnetic insulating background formed by a different band.  It is shown analytically that this system develops an instability similar to the conventional BCS 
superconducting state but mediated by magnons. However, this instability  only occurs if the Fermi velocity is larger than the magnon velocity, a property that 
can be tested experimentally. If the non-magnetic band is empty without doping, the Fermi velocity increases from zero as a function of doping. Therefore, superconductivity would only occur above a critical doping. In addition, a pseudo gap phase could appear for an anisotropic Fermi surface for intermediate densities. 

The rest of the paper is organized as follows; Section \ref{sc:model}  describes the model. Sections \ref{sc:spin} and \ref{sc:fermion} present well known derivations of the spin and fermion Hamiltonians for completeness; Section \ref{sc:interaction} discusses the spin-fermion interaction and shows how  Fr\"ohlich-like Hamiltonian\cite{frohlich50} can be obtained from the model;
Section \ref{sc:pairing} discusses of  Bardeen-Pines-like~\cite{bardin55} pairing interaction; Section \ref{sc:spingap}  describes the emergence of a spin gap for
the magnetic excitations for $q \rightarrow 0$. Section \ref{sc:case}  discusses the possibility that this model
could explain the superconducting cuprates. Finally  Section \ref{sc:summary}  summarizes. 
 
\section{Model Hamiltonian}
\label{sc:model}
\subsection{A metallic non-magnetic band in an antiferromagnetic background}
The simplest case for an antiferromagnetic insulator is a magnetic cation in an insulating lattice. The magnetic cation has, in general, 
an incomplete shell, either {\it d} or {\it f}.  The projection of the spin of the cation alternates. 
The spin excitations from the antiferromagnetic ground state are well understood.\cite{QTS,manousakis91}  

Let's consider the case of an additional electronic band centered in a different non-magnetic atom. 
Let's also assume that once the system is doped, the Fermi level lies within that band; thus, 
there is a Fermi surface that separates occupied from empty states. 

A carrier in the non-magnetic band at point $r_h$ will interact with with a spin ${\bf S_{r_\delta}}$ in the magnetic band with a Hamiltonian the form 
\begin{align}
\label{eq:hint}
H_{int}= &   \sum_{\delta}  J_2(r_\delta-r_h) {\bf S_{r_\delta}} {\bf \sigma}_{r_h} , 
\end{align} 
being
\begin{align}
J_2(\delta)= \int \frac{\phi_s(r+\delta) \phi_h(r) \phi_s(r^\prime+\delta) \phi_h(r^\prime)}{|r-r^\prime |} dr dr^\prime,
\end{align}
the direct exchange integral between the wave function associated with the localized spin $\phi_s(r) $ and a Wannier 
wave function of no-magnetic band $\phi_h(r)$. In Eq. (\ref{eq:hint}) ${\bf \sigma}_{r_h}$ is the spin operator acting on
the free carrier at $r_h$. 

The Fourier transform of Eq. (\ref{eq:hint}) leads to  Eq. (1) in Ref. \onlinecite{monthoux92}, and thus it could lead to {\it d} pairing. The difference is that, in  Ref. \onlinecite{monthoux92}, the spins that cause the spin fluctuations are the same that couple
via their own spin fluctuations. In the present case the spin system and the electronic system are separated bands which are coupled
only by the exchange interaction. Thus the spin susceptibility would be the same that in e.g. Ref. \onlinecite{sokol93}, but in this case
it would not be an approximation. Therefore, including an additional band in the picture, rather than making the problem more difficult,  avoids approximations and allows to extend the theory further (see below).

Any partially occupied  
band with overlap on the magnetic ions will become magnetically polarized unless the exchange interaction $J_2(\delta)$ of opposite
spins cancels out. In the particular case where (i) 
the non magnetic band is centered at a symmetrical position with respect to magnetic ions 
with opposite spins and  (ii) if the wave function is even under inversion with respect to that center,  $J_2(\delta)=J_2(-\delta)$ and
the exchange interaction with the opposite spins will cancel out (in first order). In that case,  a non-magnetic unpolarized state will remain stable as long as 
 the density of states at the Fermi level is lower than the Stoner criterion.
 
\subsection{Scenario where this model would apply}
Let's assume that the $d_{x^2-y^2}$ orbital has only one electron, while the rest of the $d$ orbitals have lower energy and
are fully occupied. This is the configuration commonly accepted for the Cu atoms in the superconducting cuprates.  

Note that  in the CuO$_2$ plane of a typical cuprate, the  oxygen atoms are equidistant to Cu atoms
with opposite spins.  A significant fraction of the literature argues
 that hole doping induces a partial occupation of the oxygen orbitals in the plane.~\cite{himpsel88,nucker89} 
The oxygen atom has there orbitals: $p_\sigma$ is
parallel to the direction of the Cu-Cu first neighbor direction; $p_z$ is perpendicular to the plane; and $p_\pi$ is  in the 
plane but perpendicular to the Cu-Cu direction. Most of the theoretical literature assumes that
$p_\sigma$  and $d_{x^2-y^2}$ include all the relevant physics. Therefore, this three band model~\cite{emery87} and
approximations related to it have received most of the attention.~\cite{dagotto94,imada98} 
 
\begin{figure}
\includegraphics[width=1.00\linewidth,clip=true]{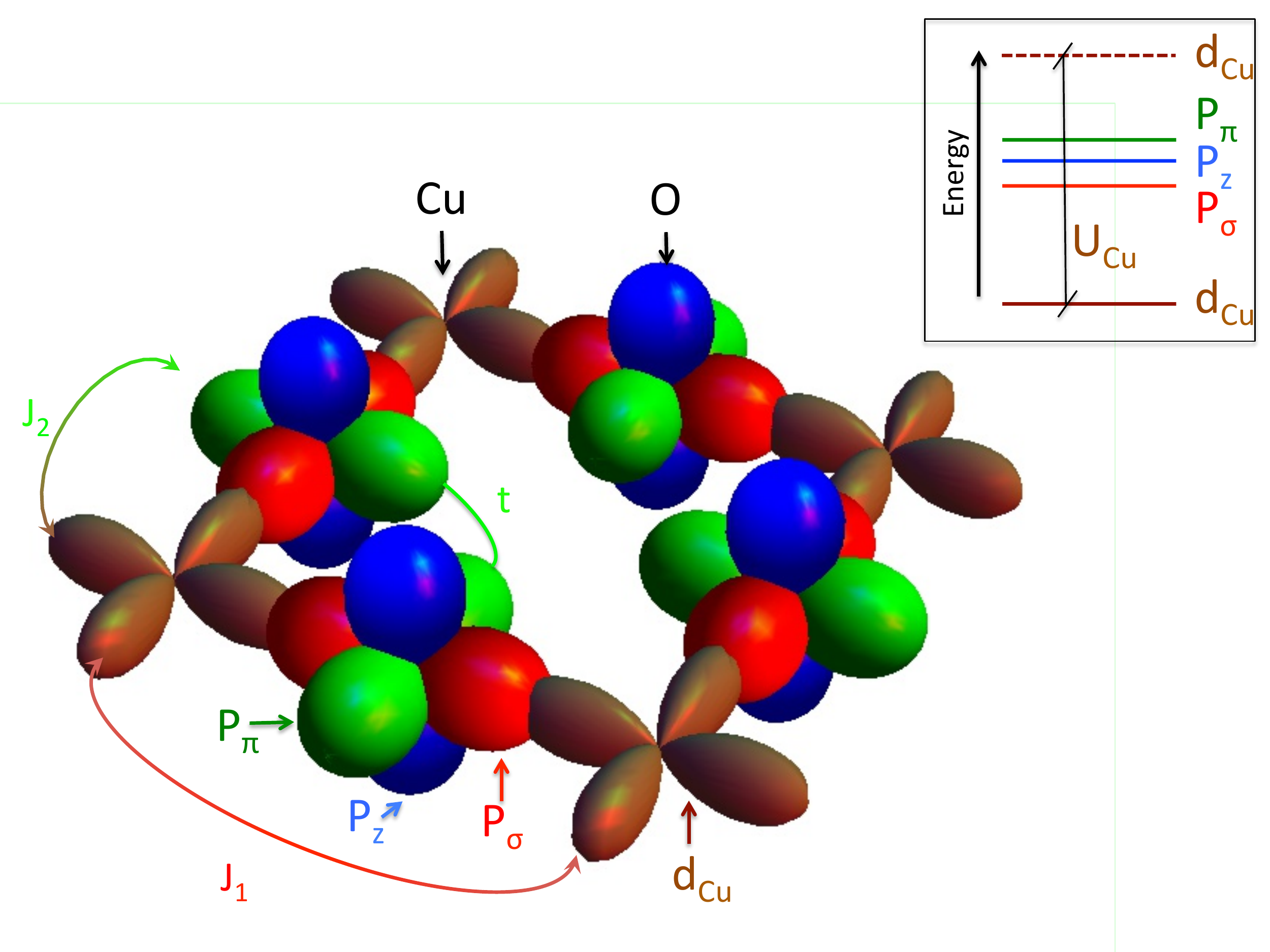}
\caption{
\label{fg:oxygen} (Color online)
Schematic representation of the energies relevant orbitals in a CuO$_2$ ignoring hopping terms and considering only
crystal field effects. The model considered in this paper 
differs from the  usual alignment of the oxygen energy levels generally accepted in the literature. If the crystal field were dominant as compared with hopping terms, the top of the conduction band would be formed by an orbital
 perpendicular to the Cu-Cu direction ($p_\pi$ in green). Since hopping terms decrease faster than crystalline fields, this arrangement will emerge for
 an expanded lattice. 
}
\end{figure}

Figure \ref{fg:oxygen} shows a schematic representation of the oxygen $p$ orbitals. Let's consider, first as an academic exercise, what
could happen if the alignment of the energy levels were the one represented in the inset. That is, what would happen if 
it were energetically favorable for the hole to occupy the $p_\pi$ orbital perpendicular to $p_\sigma$ but in the CuO$_2$ plane. 

Note that {\it d}$_{x^2-y^2}$ is even while $p_\pi$ is odd for reflections perpendicular to the plane along the Cu-Cu direction. Therefore
the hopping between $p_\pi$ and $d_{x^2-y^2}$ is zero due to symmetry. However, 
in general,  $J_2 \ne 0 $  as long as $\phi_d(r+\delta) \phi_{p_\pi}(r) \ne 0$. Therefore, in a system where the hole goes 
to $p_\pi$ one is not strictly doping the Mott insulator\cite{imada98} but a non magnetic band next to it. 

In the electron doped case~\cite{tokura89,armitage10}, this model would apply if the electrons occupy cations off the plane (e.g.  a band 
centered on the Sr ion in  the Ce$_x$ Sr$_{1-x}$CuO$_4$  system).  If the electrons go to a band centered in an orbital with ``s-like" symmetry,
the hopping integral with the $d_{x^2-y^2}$ orbital will be zero by symmetry but not the exchange. 

 It  is well established in the
literature both experimentally and theoretically\cite{manousakis91} that the ground state of the undoped  CuO$_2$ system 
 is antiferromagnetic. 
Let's analyze what could happen if all the holes go to an alternative orbital $p_\pi$ in the presence of this 
antiferromagnetic background.

The first neighbor Cu-Cu antiferromagnetic superexchange coupling $J_1$  results from the magnetic polarization of  the 
$p_\sigma$ orbital in 
between. 
In this work, it is assumed
that the hole goes to a perpendicular $p_\pi$ orbital. Therefore, in this context, the antiferromagnetic coupling $J_1$ between neighboring Cu mediated by the bridge oxygen $p_\sigma$ remains unaltered after doping.  

The rest of the paper assumes that the spin Hamiltonian is well described by a three band Hubbard model at half filling. The fermion 
Hamiltonian is formed by the $p_\pi$ and it is described by a simple tight binding model. Finally, the solution to the
spin-fermion coupling and its consequences are described. 

 \section{The spin Hamiltonian}
\label{sc:spin}
In the absence of doping, the lower energy excitations of three band Hubbard system\cite{emery87} can be modeled with a Heisenberg Hamiltonian,\cite{manousakis91,poilblanc94} though
higher energy excitations require additional terms.\cite{lorenzana99} The standard 
textbook derivation \cite{QTS} is followed here for completeness. The spin lattices are split into an ``a"  square lattice with spins pointing up with 
lattice constant $d$
and an equivalent ``b" lattice pointing down shifted  a vector {1/2,1/2}. This leads to the following Hamiltonian: 
\begin{align}
\mylabel{eq:spin}
H_s = & {J_1}  \sum_{i,j}  {\bf S^a_i} {\bf S^b_{i+j/2}} \nonumber \\  
       =  & {J_1}  \sum_{i,j} 
       \left[ \frac{1}{2}
(S^{a,+}_{i}S^{b,-}_{i+\frac{j}{2}}+S^{a,-}_{i}S^{b,+}_{i+\frac{j}{2}})+S^{a,z}_{i}S^{b,z}_{i+\frac{j}{2}}
       \right].
\end{align}
In Eq. (\ref{eq:spin}) $i$ runs over all the ``a" lattice sites and $j$ only over the first neighbors. Thus $i$ and $j$  can be one, two, three dimensional 
 vectors. In the present case $i$ and $j$ are pairs of integers while $j$ runs over four neighbors. 

The excitations of the spin Hamiltonian in Eq. (\ref{eq:spin}) can be obtained using the transformations of Holstein and Primakoff.~\cite{QTS} Approximated to lowest 
order in $1/2S$, they are given by:
\begin{align}
\mylabel{eq:holstein}
S^{a,z}_{j}  & \rightarrow S-\frac{ e^{i j.({q_1}-{q_2})d}a^\dagger_{a, q_1}a_{a, q_2}}{N_k} \\
S^{b,z}_{j}   & \rightarrow \frac{e^{-i j.({q_1}-{q_2})d}a^\dagger_{b, q_1}a_{b, q_2}}{N_k}-S  \nonumber \\
S^{a,+}_{j}   & \rightarrow \sqrt{2}e^{-i j.{q}d}\sqrt{\frac{S}{N_k}}a_{a, q} \nonumber \\
S^{a,-}_{j}    & \rightarrow \sqrt{2}e^{i j.{q}d}\sqrt{\frac{S}{N_k}}a^\dagger_{a, q} \nonumber\\
S^{b,+}_{j}  &\rightarrow \sqrt{2}e^{-i j.{q}d}\sqrt{\frac{S}{N_k}}a^\dagger_{b, q} \nonumber\\
S^{b,-}_{j}   & \rightarrow \sqrt{2}e^{i j.{q}d}\sqrt{\frac{S}{N_k}}a_{b, q}, \nonumber
\end{align}
with 
\begin{align}
[a_{k^\prime},a_k^\dagger] & = a_{k^\prime}a_k^\dagger-a_k^\dagger a_{k^\prime}=\delta_{k,k^\prime}, \\ \nonumber 
[a_{k^\prime},a_k]  &= [a^\dagger_{k^\prime},a_k^\dagger] =0 .
\end{align}
{\bf In Eqs. (\ref{eq:holstein}) and all equations that follow, obey the convention that every $q,q_1,q_2, \cdots qn $ or $k,k_1,k_2, \cdots kn $ appearing 
in the left hand side but not in the righthand side must 
be summed for every value in the Brillouin zone satisfying periodic boundary conditions}. $N_k$ is the number of $k$-points satisfying periodic 
boundary conditions for a given supercell.  Replacing Eqs. (\ref{eq:holstein}) into Eq. (\ref{eq:spin}) and removing two-body terms, one obtains:
 \begin{align}
 \mylabel{eq:fourier}
H_s \simeq  &{J_1} z S
\left[ 
a^\dagger_{a, q} a_{a, q}+ a^\dagger_{b, q} a_{b, q}+ a_{a, q} a_{b, q} \gamma ({q}) 
 + a^\dagger_{b, q} a^\dagger_{a, q} \gamma ({q}) \right]-{J_1} N_k z S^2,
 \end{align}
 $z$ being the number of first neighbors and  
 \begin{align}
 \gamma ({q})= \frac{1}{z}\sum_j e^{{\bf i} j\cdot q \frac{d}{2}}.
 \end{align}
 
One can use a transformation due to Bovoliubov defined by
 \begin{align}
 \mylabel{eq:bovoliubov}
a_{b, q} \rightarrow  & a_{\beta, q} u(q)+a^\dagger_{\alpha, q} v(q) \\
a_{a ,q} \rightarrow  & a_{\alpha ,q} u(q)+  a^\dagger_{\beta, q} v(q) \nonumber \\
a^\dagger_{{b}, {q}}    \rightarrow & a^\dagger_{\beta ,q} u(q)+a_{\alpha, q} v(q) \nonumber \\
a^\dagger_{{a}, {q}} \rightarrow &a^\dagger_{\alpha, q} u(q)+a_{\beta, q} v(q) \nonumber,
\end{align}
 with
 \begin{align}
 &u({q})^2-v({q})^2 =1 \; \text{ and } \\ \nonumber 
  &u({q})^2+v({q})^2+2 u({q}) v({q}) \gamma ({q})=0.
  \end{align}
  
 Replacing Eqs. (\ref{eq:bovoliubov}) into Eq.  (\ref{eq:fourier}) one obtains
 \begin{align}
 \mylabel{eq:h0}
 H_s= \omega_{\alpha, q}  a^\dagger_{\alpha, q}  a_{\alpha, q}+ \omega_{\beta, q}  a^\dagger_{\beta, q}  a_{\beta, q}-{J_1} N_k z S^2,
 \end{align}
 with
 \begin{align}
 \mylabel{eq:omega}
  \omega_{\alpha, q}=
   \omega_{\beta, q}  =
 {J_1} S \left[
 1+2 v({q})^2+2 v({q}) \gamma ({q}) \sqrt{v({q})^2+1} 
\right].
 \end{align}
 
 While only the case of an antiferromagnetic background has been considered in this work, a  long range antiferromagnetic ordering is  not strictly required. Only a background with antiferromagnetic-like excitations that go  to zero
linearly with $|q|$ is required.

 \section{The fermion Hamiltonian}
 \label{sc:fermion}

Let's consider the limit case in which all the holes introduced by doping go the oxygen $p_\pi$ orbitals in the Cu0$_2$ plane but perpendicular to the
O-Cu-O bonds. 
 These orbitals 
 have identical hoppings $t$ to the four for first neighbor perpendicular $p_\pi$ orbitals. One can also formulate this problem as two interpenetrated lattices
 of $p_x$ and $p_y$ with only hoppings to each other. Since the cell size is doubled by the antiferromagnetic ordering and O buckling, there
 are four orbitals per cell. The oxygens  follow a smaller square mesh. Thus, the eigenenergies to first approximation, can be obtained folding the eigenvalues
 of a square lattice with hoppings to first neighbors given by
\begin{align}
\mylabel{eq:Hf}
H_f= \epsilon(\uparrow, k) 
c^\dagger_{\uparrow,k}c_{\uparrow,k}+\epsilon(\downarrow, k)  c^\dagger_{\downarrow,k}c_{\downarrow,k}
,
\end{align}
 with
 \begin{align}
 \epsilon({\bf k})= 2t \left[ \cos\left( \frac{d}{2} k_x \right)+\cos\left({\frac{d}{2}} k_y  \right) \right],
 \end{align}
 and 
 \begin{align}
 c^\dagger_{k}c_{k^\prime}+c_{k^\prime}c^\dagger_{k} & =\delta_{k,k^\prime} \\ \nonumber
 c_{k} c_{k^\prime} = c^\dagger_{k} c^\dagger_{k^\prime} & =0.
 \end{align}
 Note that the lattice vectors connect two Cu with the same spin and are oriented in the $(1,\pm 1,0)$ directions of the crystal.
 Second neighbor hoppings between parallel $p_\pi$ orbitals could be included in the model. Second neighbor hoppings would make the 
 Fermi surface anisotropic. This anisotropy will in turn affect the superconducting state. 
 
 \section{The spin-fermion interaction}
 \label{sc:interaction}
 Let's consider what happens when a hole sits in the  $p_\pi$ orbital.
 The unpaired spin left in $p_\pi$ will couple with the unpaired spins of the neighboring {\it d} orbitals in copper as
 \begin{align}
 \mylabel{eq:interaction}
 H_{s-f} = & J_2  \sum_{i,j} \left( {\bf S^a_i} +{\bf S^b_{i+j/2}} \right) {\bf \sigma}_{i,j/4}  = \nonumber \\  
 \frac{1}{2} J_2 \sum_{i,j} \nonumber
 &\left[ 
 c^\dagger_{{\uparrow}, i+\frac{1}{4} j} c_{{\downarrow}, i+\frac{1}{4} j}(S^{a,-}_{i}+S^{b,-}_{i+\frac{1}{2} j})+\ c^\dagger_{{\downarrow} ,i+\frac{1}{4} j} c_{{\uparrow}, i+\frac{1}{4} j}(S^{a,+}_{i}+S^{b,+}_{i+\frac{1}{2} j}) +  \right.  \\ 
  & \left. + (S^{a,z}_{i}+S^{b,z}_{i+\frac{1}{2} j})(c^\dagger_{{\uparrow} ,i+\frac{1}{4} j} c_{{\uparrow} ,i+\frac{1}{4} j}-c^\dagger_{{\downarrow} ,i+\frac{1}{4} j} c_{{\downarrow} ,i+\frac{1}{4} j})
 \right],
 \end{align} 
 where $i$ and $j$ follow the same convention as in Eq. (\ref{eq:spin}).  In the case of electron doping
 assuming that the electron goes to a cation other than copper (e.g. Nd in the NdCuO$_4$ system) the
 interaction Hamiltonian will be different but the differences to the derivation that follows will be minimal.

 Replacing Eqs. (\ref{eq:holstein}) in Eq.(\ref{eq:interaction}) doing a Fourier transform over 
 the fermions and summing over all $i$ one obtains
 \begin{align}
 \mylabel{eq:spinfer}
 H_{s-f} \simeq
 &\frac{ {J_2}}{{2 N_k}}
\left\{
 \sqrt{2 S N_k}\;  \gamma\left(\frac{{q_1}}{2}\right) 
 \left[
 c^\dagger_{\uparrow,{k_1}-{q_1}} c_{\downarrow {k_1}}
   \left(a_{{b},{q_1}}+a^\dagger_{{a},{q_1}}\right)+ \right. \right. \\ \nonumber
  & \left.\;\;\;\;\;\;\;\;\;\;\;\;\;\;\;\;\;\;\;\;\;\;\;\;\;\;\;\;\;\;\;\;\;\;\;\;\;
  +c^\dagger_{\downarrow {k_1}+{q_1}} c_{\uparrow,{k_1}}\left( a_{{a},{q_1}}+a^\dagger_{{b},{q_1}}\right)
 \right]+ \\  \nonumber
 &+ \gamma\left(\frac{{q_1}-{q_2}}{2}\right) 
 \left[ 
 a^\dagger_{{a},{q_2}} a_{{a},{q_1}} 
 \left(c^\dagger_{\downarrow, {k_1}+{q_1}-{q_2}} c_{\downarrow, {k_1}}-c^\dagger_{\uparrow, {k_1}+{q_1}-{q_2}} c_{\uparrow,{k_1}}\right) 
 \right. 
 + \\ \nonumber
 & \;\;\;\;\;\;\;\;\;\;\;\;\;\;\;\;\;\;\;\;\;\;
+ \left.  \left.  a^\dagger_{{b},{q_2}} a_{{b},{q_1}}
\left(c^\dagger_{\uparrow, {k_1}+{q_2}-{q_1}} c_{\uparrow ,{k_1}}-c^\dagger_{\downarrow, {k_1}+{q_2}-{q_1}} c_{\downarrow, {k_1}}\right) \right]
 \right\}.
 \end{align}
 Note that the first two lines in Eq. (\ref{eq:spinfer}) have a structure that resembles the Fr\"ohlich Hamiltonian. These lines can be
 traced back to  the $S^{+}\sigma^{-}+S^{-}\sigma^{+}$ terms in the spin-fermion coupling in Eq. (\ref{eq:interaction}). The remaining
 two lines are absent in the Fr\"ohlich Hamiltonian and result from the product $S^{z}\sigma^{z}$. Therefore, this spin-fermion model interaction
 gives rise to magnon absorption, emission and Raman-like scattering terms. 
 
 Replacing Eq. (\ref{eq:bovoliubov}) into Eq. (\ref{eq:spinfer}) one obtains
\begin{align}
\mylabel{eq:converthint}
H_{s-f}  &\simeq
J_2 \nonumber
\left\{  \
\sqrt{\frac{S}{{2 N_k}}} \gamma \left(\frac{{q_1}}{2}\right)
U_p(q_1)
\left[
c^\dagger_{\uparrow,{k_1}-{q_1}} c_{\downarrow,{k_1}} ( a_{\beta,{q_1}} +a^\dagger_{\alpha,{q_1}} )+c^\dagger_{\downarrow,{k_1}+{q_1}} c_{\uparrow,{k_1}} (a_{\alpha,{q_1}}+a^\dagger_{\beta,{q_1}}) 
\right]+ 
\right. 
\\ & \nonumber
 \frac{ \gamma \left(\frac{{q_1}-{q_2}}{2}\right)}{{2 N_k}} 
 \left[
 U_u(q_1,q_2)
(c^\dagger_{\downarrow,{k_1}+{q_1}-{q_2}}  c_{\downarrow,{k_1}}- c^\dagger_{\uparrow,{k_1}+{q_1}-{q_2}}  c_{\uparrow,{k_1}}) 
(a^\dagger_{\alpha,{q_2}} a_{\alpha,{q_1}}  -a^\dagger_{\beta,{q_1}} a_{\beta,{q_2}})
+ 
 \right. 
 \\  &  
\left.\left.
\;\;\;\;\;\;\;\;\;\;\;\;\;\;\;\;\;U_v(q_1,q_2)
(c^\dagger_{\uparrow,{k_1}+{q_2}-{q_1}} c_{\uparrow,{k_1}}-c^\dagger_{\downarrow,{k_1}+{q_2}-{q_1}} c_{\downarrow,{k_1}})(a_{\alpha,{q_2}} a_{\beta,{q_1}}-a^\dagger_{\beta,{q_2}} a^\dagger_{\alpha,{q_1}}) 
\right] \right\},
\end{align}
with 
\begin{align}
\label{eq:Up}
U_p(q_1)= & u({q_1}) +v({q_1})  \; , \\
U_u(q_1,q_2) = & 
u({q_1}) u({q_2})- v({q_1}) v({q_2}) \;,\text{ and } \\
  U_v(q_1,q_2) = &
u({q_1}) v({q_2}) - u({q_2}) v({q_1}) .
\end{align}
Note again, in the first line in Eq. (\ref{eq:converthint}),  an structure that resembles the Fr\"ohlich Hamiltonian. The main differences are
that (i) the factor that multiplies the operators is real and (ii) the absorption or emission of a magnon flips the spin of the fermion.  

\section{The pairing interaction}
\label{sc:pairing}
The complete Hamiltonian is then given by
\begin{align}
\label{eq:h}
H= & H_s+H_f+H_{s-f} \nonumber \\
  = & H_0+H_{s-f}.
\end{align}
The eigenvalues of $H$ are not changed by linear transformations of  form
\begin{align}
\tilde{H}= & e^{P}He^{-P}.  \\
\end{align}
 Using the Baker-Campbell-Hausdorf formula 
\begin{align}
\mylabel{eq:expan}
\tilde{H} \simeq & H + [P, H] + 1/2 [P, [P, H]] + 1/3! [P, [P, [P, H]]]+ \cdots ,
\end{align}
being $[P,H]= PH-HP $. 
Let's define $P$ as  
\begin{align}
\label{eq:s}
P=J_2 P_1+  J_2^2 P_2.
\end{align}
Replacing Eq.(\ref{eq:h}) and Eq.(\ref{eq:s}) into Eq.(\ref{eq:expan}) and choosing the terms linear in $J_2$ and enforcing 
them to be zero\cite{QTS} one obtains
\begin{align}
J_2 [P_1, H_0] = -H_{s-f} .
\end{align}
Note that, if $h$ is a product of $a^\dagger_{a,q}$, $a_{a,q}$, $c_{\sigma,k}^\dagger$, and $c_{\sigma,k}$ , the commutator $[h,H_0]$  is a scalar given by a sum  
of eigenvalues  of $H_0$. Therefore, any term $h_1$ in $H_{s_f}$ with $[h_1,H_0] \ne 0 $ can be removed 
from $\tilde{H}$ to first order in $J_2$ using the 
transformation in Eq. (\ref{eq:expan}) and including in $P_1$  a 
term of the form $ J_2 h_1/[h_1,H_0] $. 

Note that when $q_1=q_2$,  some terms $h$ in the second line of Eq. (\ref{eq:converthint}) give 
$[h,H_0]$=0, and cannot be removed. 
Any term $h$ with $[h,H_0]=0$ applied to the an eigenstate  of $H_0$, returns the same eigenstate times a factor. 
These factors are the correction to the energy of the ground state in first perturbation theory.  

The expansion of Eq.  (\ref{eq:converthint}) gives 12 different terms to be considered in $P_1$. The expression of $P_1$ is given
in Appendix \ref{ap:p1}.

Similarly, retaining the terms of order $J_2^2$  and enforcing them to be zero one obtains:
\begin{align}
\label{eq:s2}
[P_2, H_0]= -1/2 [P_1, H_{s-f]}]  \;,
\end{align}
which implies that for any term $h_2$ in $1/2 [P_1, H_{s-f}]$ which does not commute with $H_0$ one can add a term 
to $P_2$ that removes it from $\tilde{H}$ up to second order in $J_2$. Including all the terms $h_2$ that do not commute with 
$H_0$ in $P_2$ will yield only the second order correction in $J_2$ to the ground state energy of $H_0$.  
However, each one of the terms with $[h_2,H_0]\ne 0 $ can induce an instability 
in the ground state of the spins, the holes or both analogous to the one discussed by BCS. 

This study focuses on an instability towards a BCS-like superconducting ground state.
Thus,  only the terms $h_2$ that have a structure similar to the Bardeen-Pines Hamiltonian\cite{bardin55} and the ones that 
commute with $H_0$ are  excluded from $P_2$ and retained in $\tilde{H}$. After rather laborious calculations and assuming that 
$\epsilon(\uparrow,k)=\epsilon(\downarrow,k)= \epsilon(k)$ and $\omega(\alpha,q)=\omega(\beta,q)= \omega(q)$ one obtains. 
\begin{align}
\mylabel{eq:HJ2}
\tilde{H}= & H_0+ 1/2  [P_1, H_{s-f}] \\ \nonumber 
	= & H_0 +E_1+E_2 + 
c^\dagger_{\downarrow,{k_2}+{q_1}} c^\dagger_{\uparrow,{k_1}-{q_1}} c_{\uparrow,{k_2}} c_{\downarrow,{k_1}}  \times \\ \nonumber &
\times \left\{  - V_0(k_1,q_1) \right.
+ 
 V_{a^\dagger, a} (k_1,k_2,q_1)
 \left( a^\dagger_{\alpha ,{q_1}} a_{\alpha ,{q_1}}+a^\dagger_{\beta ,{q_1}} a_{\beta ,{q_1}} \right) + 
\\ \nonumber &
\left.
V_{a^\dagger ,a^\dagger}(k_1,k_2,q_1)   a^\dagger_{\alpha ,{q_1}}  a^\dagger_{\beta ,{q_1}}    
 + V_{a, a}(k_1,k_2,q_1)   a_{\alpha ,{q_1}}  a_{\beta ,{q_1}} 
  \right\}
  +\\  \nonumber - & [P_2,H_0],
\end{align}
with 
\begin{align}
V_0(k_1,q_1) &= \frac{ S J_2^2  {U_p}({q_1})^2 \gamma \left(\frac{{q_1}}{2}\right)^2}{N_k}  g(k_1,q_1) \text {, with }\\ \nonumber
g(k_1,q_1) & =
\left[
\frac{ \omega (q_1)}{ (\epsilon (k_1\!-\!q_1)\!-\!\epsilon (k_1))^2
\!-\! \omega (q_1)^2} \!+\! 
\frac{ \omega (q_1)}{ (\epsilon (k_1\!+\!q_1)\!-\!\epsilon (k_1))^2
\!-\!\omega (q_1)^2}
\right],  
\end{align}
and
\begin{align}
\mylabel{eq:Vada}
 & V_{a^\dagger ,a}(k_1,k_2,q_1) =  
\frac{1}{8} {Jz_2}^2 \frac{1}{N_k^2} \gamma \left(\frac{{q_1}}{2}\right)^2 \times \\ \nonumber
&\left\{ U_u({q_2},{q_2}\!-\!{q_1})^2\left[\frac{1}{\!-\!\epsilon ({k_2}\!+\!{q_1})\!+\!\epsilon ({k_2})\!+\!\omega ({q_2}\!-\!{q_1})\!-\!\omega ({q_2})}\!-\!\frac{1}{\!-\!\epsilon ({k_1}\!-\!{q_1})\!+\!\epsilon ({k_1})\!-\!\omega ({q_2}\!-\!{q_1})\!+\!\omega ({q_2})}\right]\!+\! \right. \\ \nonumber 
&+ 
U_v({q_2},{q_2}\!-\!{q_1})^2\left[\frac{1}{-\epsilon ({k_1}\!-\!{q_1})\!+\!\epsilon ({k_1})\!+\!\omega ({q_2}\!-\!{q_1})\!+\!\omega ({q_2})}\!-\!\frac{1}{-\epsilon ({k_2}\!+\!{q_1})\!+\!\epsilon ({k_2})\!-\!\omega ({q_2}\!-\!{q_1})\!-\!\omega ({q_2})}\right]\!+\! \\ \nonumber 
&+  U_u({q_2},{q_1}\!+\!{q_2})^2\left[\frac{1}{-\epsilon ({k_1}\!-\!{q_1})\!+\!\epsilon ({k_1})\!+\!\omega ({q_1}\!+\!{q_2})\!-\!\omega ({q_2})}\!-\!\frac{1}{-\epsilon ({k_2}\!+\!{q_1})\!+\!\epsilon ({k_2})\!-\!\omega ({q_1}\!+\!{q_2})\!+\!\omega ({q_2})}\right]\!+\! \\ \nonumber 
&+ 
\left.
U_v({q_2},{q_1}\!+\!{q_2})^2\left[\frac{1}{-\epsilon ({k_2}\!+\!{q_1})\!+\!\epsilon ({k_2})\!+\!\omega ({q_1}\!+\!{q_2})\!+\!\omega ({q_2})}\!-\!\frac{1}{-\epsilon ({k_1}\!-\!{q_1})\!+\!\epsilon ({k_1})\!-\!\omega ({q_1}\!+\!{q_2})\!-\!\omega ({q_2})}\right] \right\}.
 \end{align}
 
The expresions for $V_{a^\dagger, a^\dagger}$ and $V_{a, a}$ are presented in Appendix \ref{ap:adad}. 
  
In Eq. (\ref{eq:HJ2}) $E_1$ and $E_2$ are respectively the first order and second order corrections in $J_2$ to the ground state of $H_0$ under the perturbation  $H_{s-f}$. In (\ref{eq:HJ2})  $P_2$ contains $98$ terms of order $J_2^2$. While the expression for $P_2$ has been handled with
a package developed by the author in the context of the symbolic algebra Mathematica program, it is clearly impractical to reproduce it in printed form due to its size.

This work  assumes that all the terms included in the quasiparticles with the transformation  $e^{-P_2}H e^{P_2}$ do not soften a mode towards a different possible instability (e.g. ferromagnetic, spin-density wave, charge separation etc.). In principle, this system could present
other instabilities that compete with a BCS-like instability, but their study is beyond the scope of this paper. 

\section{The superconducting state}
\label{sc:HTc}
\subsection{First necessary condition for BCS-like superconductivity}
For a BCS-like superconductivity to occur, the BCS-like ground state must have lower energy than any other possible wave function. 
A necessary condition is that the BCS-like solution must have lower energy than the normal metallic ground state.

Let's consider first the term proportional to $V_0(k_1,q_1)$ in Eq. (\ref{eq:HJ2}). Note in $g(k_1,q_1)$ the change of sign as compared with the 
electron-electron interaction obtained\cite{bardin55} from the Fr\"ohlich 
Hamiltonian.~\cite{frohlich50} This change of sign can be traced to a missing imaginary constant.   $H_{s-f}$ is real while the Fr\"ohlich Hamiltonian is imaginary.  
 For the case of $H_{s-f}$, an attractive interaction  occurs  if $\omega (q_1) ^2 < (\epsilon (k_1-q_1)-\epsilon (k_1))^2$ and 
 $\omega (q_1) ^2 < (\epsilon (k_1+q_1)-\epsilon (k_1))^2$. Negative terms will appear if the fermion band width is larger than the spin excitation band width. 
 For superconductivity to occur, a gap must appear for the low energy excitations. This  implies that negative terms must be present for 
 $q_1 \rightarrow 0$, which involve excitations of electron hole pairs  near the Fermi surface.
 Therefore, a necessary condition for an attractive term near the Fermi surface is then given by
 \begin{align}
 \mylabel{eq:condsup}
 J_1 S d < \left| \nabla_k \epsilon({k_F})  \right|,
 \end{align}
being $J_1 S d$ the speed of the antiferromagnetic magnetic excitations  for $q\rightarrow 0$,  
$\nabla_k \epsilon({k_F})$ the gradient of the fermion band at the Fermi surface and  $|\nabla_k \epsilon({k_F})|$
the Fermi velocity. Assuming that, (i) this condition is the only one limiting superconductivity, (ii) all the holes go to the $p_\pi$ orbitals, (iii)  
doping is 100\% effective, (iv) the oversimplified tight binding in Eq. (\ref{eq:Hf}) is valid (v)
 the critical doping is ~2.5\%,  and (vi) $J_1= 0.1$eV, then the
condition in Eq. (\ref{eq:condsup}) would be satisfied if $t\approx 0.5$ eV, that is a band width of $4$ eV. Higher values 
would be obtained if one assumes that doping is not efficient or that carriers remain in the reservoir.   

Note that repulsive terms will appear when $|\cos(\phi_\kappa)| |\nabla (\epsilon(k_F))| < J_1 S d $ with
$\cos(\phi_\kappa)= q.\nabla (\epsilon(k_F)/|q|/|\nabla (\epsilon(k_F))|$.
These repulsive terms do not contribute to 
an instability and thus can they be added to $P_2$ as 
\begin{align}
P_2^{\prime}= \theta[-g(k_1,q_2)]  \frac{ V_0(k_1,q_1)  c^\dagger_{\downarrow,{k_2}+{q_1}} c^\dagger_{\uparrow,{k_1}-{q_1}} c_{\uparrow,{k_2}} c_{\downarrow,{k_1}}}
{\epsilon(k_1)+\epsilon(k_2)-\epsilon(k_1-q_1)-\epsilon(k_2+q_2)} ,
\end{align}
being $\theta(x)$ the heaviside step function.  The resulting effective Hamiltonian only retains the negative terms.

\subsection{BCS-like superconductivity}

Taking into account the fermion Hamiltonian Eq.(\ref{eq:Hf}) and the term proportional to $V_0$ in Eq. (\ref{eq:HJ2}), choosing $k_2=k_1=k$ and $k_1-q_1=k^\prime$ one arrives
to a reduced Hamiltonian of the form 
\begin{align}
\mylabel{eq:Hred}
H_{red}=
\epsilon(k)\left[c^\dagger_{\uparrow,k}c_{\uparrow,k}+ c^\dagger_{\downarrow,k}c_{\downarrow,k} \right]-
V_0(k,q)  \theta[g(k,q)]
c^\dagger_{\downarrow,-k-q} c^\dagger_{\uparrow,k+q} c_{\uparrow,-k} c_{\downarrow,{k}},  
\end{align}
 which has an almost identical structure to the one find in textbooks\cite{QTS} from which superconductivity 
 can be derived. The main difference is the factor $\theta[g[k,q] ]$  which appears when the 
 repulsive terms are included in $P_2$. 
 
 It is interesting that the absorption and emission  of a mangnon that flips spins for small $q$ at opposite sides of the 
Fermi surface is equivalent to the absorption and emission of a phonon with $q\simeq 2 k_F$, which does not flip spins.

It is long known that the reduced Hamiltonian in Eq. (\ref{eq:Hred}) gives rise to a state with lower energy than the metallic 
ground state (which is given by $|\Psi \rangle =\prod_k^{<k_F} c^\dagger_{\uparrow,k}c^\dagger_{\downarrow,k} |\Phi_{vac} \rangle$). BCS chose a step function for $V_0(k,q)$  and a ground state wave function of the form 
\begin{align}
\mylabel{eq:BCS}
|\Phi _{BCS} \rangle =\prod_k (u_k+ v_k c^\dagger_{\downarrow,-k} c^\dagger_{\uparrow,k}  )| \Phi_{vac} \rangle.
\end{align}
One thus can assume that the ground state $|\Phi \rangle$ is BCS-like:  if has a form similar to $\Phi_{BCS}$ but with 
more structure due to the effective interaction: 
\begin{align}
\mylabel{eq:phi}
| \Phi \rangle = \theta[g(k,q)] n(\uparrow,-k) n(\downarrow,k)[1-n(\uparrow,k+q)][1-n(\downarrow,-k-q)]
| \Phi_{BCS} \rangle,
\end{align}
with $n(\sigma,k)= c^\dagger_{\sigma,k} c_{\sigma,k}$.
Taking the average of the effective interaction in Eq. (\ref{eq:Hred}) one obtains
\begin{align}
\mylabel{eq:delta}
\Delta_{\Phi}=\Delta_{H_0}
-V_0(k,q)   \eta(\Phi,k,q).
\end{align}
with 
 \begin{align}
 \label{eq:eta}
 \eta(\Phi,k,q) = & \frac{ \langle \Phi | c^\dagger_{\downarrow,-k-q} c^\dagger_{\uparrow,k+q} c_{\uparrow,-k} c_{\downarrow,{k}}    | \Phi \rangle} {\langle \Phi | \Phi \rangle}\theta[g(k,q)] \; .
 \end{align}

Note in Eqs. (\ref{eq:phi}), ({\ref{eq:delta}}), and (\ref{eq:eta}) that $|\Phi \rangle$, by construction, selects the negative terms in the effective interaction. 
 In Eq. (\ref{eq:delta}) $\Delta_{H_0}$ is the difference of average energy of $H_0$ in the BCS-like state and the normal metallic state. Since the metallic state is the lowest energy configuration for $H_0$, $\Delta_{H_0}$ must be larger than zero. Thus,
 $\eta(\Phi,k,q) >0 $ in order to obtain $\Delta_{\Phi} < 0$. 
Equation (\ref{eq:phi}) enforces $\eta(\Phi,k,q) = 0$ whenever there is no energy gain in the effective interaction, since keeping $\eta(\Phi,k,q) \ne 0$ implies an energy cost in 
 $\Delta_{H_0}$. 
 
\section{ Renormalization of the magnon frequencies for $q\rightarrow 0$}
 \label{sc:spingap}
A significant difference with the Bardeen-Pines interaction deduced from the Fr\"ohlich Hamiltonian is the appearance of
corrections to the magnon frequencies [see terms involving $V_{a^\dagger,a}(k_1,k_2,q_1,)$, $V_{a^\dagger,a^\dagger}(k_1,k_2,q_1,)$, and  
$V_{a,a}(k_1,k_2,q_1)$  in Eq. (\ref{eq:HJ2})].

 In the case of the electron-phonon interaction, the effective Hamiltonian decouples
electrons and phonons to second order in the coupling constant. For the Fr\"ohlich Hamiltonian case, phonon-self energies are renormalized including 
fourth order terms in the expansion Eq. (\ref{eq:expan})] .  
Thus, the fact that magnons are renormalized already at second order in $J_2$, is a significant difference 
with the phonon case.

In addition, for the present model, a number of other operators appear in second order in $J_2$ which were removed from the effective Hamiltonian in Eq. ( \ref{eq:HJ2})  and \ included in $P_2$. Though the contribution to the total energy of all those terms can be handled
as a perturbation, a mayor  concern that remains is that they could contribute to scattering and dissipation of the superconducting state. 
That would not occur, however, if there is a gap for the spin excitations for $q \rightarrow 0$ induced by the superconducting state.    

In order to study the renormalization of the magnon frequencies for $q \rightarrow 0$  one must consider the remaining terms in the effective Hamiltonian in Eq. (\ref{eq:HJ2}) which are: 
\begin{align}
H_{mg}  & = \omega_{ q} \left( a^\dagger_{\alpha, q}  a_{\alpha, q}+ a^\dagger_{\beta, q}  a_{\beta, q} \right) 
  +  
c^\dagger_{\downarrow,{k_2}+{q}} c^\dagger_{\uparrow,{k_1}-{q}} c_{\uparrow,{k_2}} c_{\downarrow,{k_1}}  \times
 \\ \nonumber &
\times
\left[ V_{a^\dagger, a} (k_1,k_2,q)
 \left( a^\dagger_{\alpha ,{q}} a_{\alpha ,{q}}+a^\dagger_{\beta ,{q}} a_{\beta ,{q}} \right) + 
V_{a^\dagger ,a^\dagger}(k_1,k_2,q)   a^\dagger_{\alpha ,{q}}  a^\dagger_{\beta ,{q}}     + V_{a, a}(k_1,k_2,q)   a_{\alpha ,{q}}  a_{\beta ,{q}} 
 \right]. 
\end{align}
Taking the  average over the fermionic wave function of the  BCS-like ground state one obtains. 
\begin{align}
\frac{\langle \Phi |H_{mg} | \Phi \rangle}  {\langle \Phi | \Phi \rangle}  & =
\left[ \omega_{ q}+\eta(\Phi,k,q) V_{a^\dagger, a} (k,-k,q) \right] \left( a^\dagger_{\alpha, q}  a_{\alpha, q}+ a^\dagger_{\beta, q}  a_{\beta, q} \right) 
 +  
 \\ \nonumber &
\eta(\Phi,k,q)
\left[ 
V_{a^\dagger ,a^\dagger}(k,-k,q)   a^\dagger_{\alpha ,{q}}  a^\dagger_{\beta ,{q}}     + V_{a, a}(k,-k,q)   a_{\alpha ,{q}}  a_{\beta ,{q}} 
 \right], 
\end{align}
with 
$\eta(\Phi,k,q) \ge 0$ i for the BSC-like ground state $| \Phi \rangle$ [see Eq. (\ref{eq:delta}) and related
discussion]. 
If the average, instead, is done over a normal metallic fermionic ground state 
$|\Psi \rangle $, one  obtains 
$\eta(\Psi,k,q)=0$, which implies that the shifts to the magnon frequencies appear with a BCS-like ground state. This
state would occur only if   Eq. (\ref{eq:condsup}) is satisfied in part of the Fermi surface. Thus a shift on the 
magnon frequencies would occur in the pseudo gap phase or in the superconducting phase 
but is not present for a normal metal. Therefore, within the context of this model, any shift in the spin excitations 
contains information of the structure of the superconducting electronic state. 

\subsubsection{
The spin gap
}
 Since the effect of the terms 
involving $a^\dagger a^\dagger$ and $a a$ can be absorbed in a transformation, the renormalized frequencies 
are given by 
\begin{align}
\mylabel{eq:shift}
\tilde{\omega}_{q}=\omega_{ q}+\eta(\Phi,k,q) V_{a^\dagger, a} (k,-k,q). 
\end{align}

In order to estimate the renormalization of the frequency,  one must find an approximation for  
$V_{a^\dagger, a} (\phi,k,-k,q \rightarrow 0)$ using Eq. (\ref{eq:Vada}).
 
Assuming a conical form for $\omega(q)= |q| S J_1 $, one can write
\begin{align}
\mylabel{eq:zeta}
\omega(q+q_1) & =  \zeta(q,q_1) \left[ \omega(q)+\omega(q_1) \right] \\ \nonumber
	\text{with } & \zeta(q,q_1) = \sqrt{ 1-\frac { 2 \left( |q| |q_1|-q \cdot q_1 \right )}{\left|q+q_1\right|^2} } .
\end{align}
being $\zeta(q,q_1)$ in the interval $[0,1]$ for every $q$ and $q_1$. 

Since in the limit $q\rightarrow 0 $, $U_u(q_1,q_1\pm q) \rightarrow 1 $,
$U_v(q_1,q_1\pm q) \rightarrow 0 $, one obtains  

\begin{align}
\label{eq:step}
V_{a^\dagger,a}(\Phi,k,-k,q \rightarrow 0)=\frac{1}{4} J_2^2 \frac{1}{N_k^2} \gamma \left(\frac{{q}}{2}\right)^2
\sum_{k,q_1} \left[
\frac{ {\omega_{\zeta-}} }{\omega_\kappa ^2-\omega_{\zeta-}^2}
+\frac{ {\omega_{\zeta+}} }{\omega_\kappa ^2-\omega_{\zeta+}^2} \right],
\end{align}
with 
\begin{align}
\omega_{\zeta\pm} = \left[ \omega(q)+\omega(q_1)
\right] \zeta(\pm q,q_1),
\end{align}
and
\begin{align}
\omega_\kappa = & \epsilon(k+q)-\epsilon(k)+\omega(q_1). \\ \nonumber
\simeq & \nabla \epsilon(k).q+\omega(q_1).
\end{align}
The sums over $k$ and $q_1$ must be done only in Eq. (\ref{eq:step}) but there are not implicit sums in the next two. 
Using $$\zeta(q,q_1) \rightarrow 1+ \frac{\omega(q) [\cos(\phi)-1]}{\omega(q_1)} $$
with $\cos(\phi)= q.q_1/ (|q| |q_1|),$ and $q_1>q$ done obtains
\begin{align}
\mylabel{eq:limit}
V_{a^\dagger,a}(\Phi,k,-k,q \rightarrow 0)=
\frac{J_2^2 \left[J_1 S d  + |q_1| |\nabla \epsilon(k)| \cos(\phi_\kappa) \right] }{4 N_k^2 |q_1| 
\left[|\nabla \epsilon(k)|^2 \cos[\phi_k]^2-(J_1 S d) ^2 \cos[\phi ]^2\right]
}.
\end{align}
Provided that Eq. (\ref{eq:condsup}) is satisfied (see related discussion), $\eta(k,-k,q) \ne 0$ only when the denominator 
of Eq. (\ref{eq:limit}) is positive. There is an implicit sum in Eq. (\ref{eq:shift}) over all $k$ that cancels out 
the contribution of the term involving $\cos(\phi_\kappa)$ in the numerator.

Replacing Eq. (\ref{eq:limit}) into Eq. (\ref{eq:shift}) one obtains a positive shift for $\omega(q\rightarrow 0)$ that creates a spin gap. The spin gap prevents inelastic scattering of the superconducting state with the spin excitations below a critical current.
 
 Note in Eq. (\ref{eq:limit}) that the contribution to the shift in the spin excitations diverges as $q_1 \rightarrow 0$. This divergence is a result of the second order perturbation approach followed here which is not correct for nearly degenerate energies.
 In practice, a matrix must be solved when the energy spacing between different eigenstates in $H_0$ is smaller than $|J_2|$. The contribution to the shift of each term is therefore limited by $J_2$. The most likely scenario is that a conical dispersion for $\omega(q)$ is
 replaced by a function in which the lower energy excitations are shifted more than the higher energy ones  
 (e.g. $\tilde{\omega}(q)= J_1 S z \frac{q^2}{\delta+|q|} $ ).
 
 Note also that the shift in $\omega(q)$ is dependent on the direction of $q$ 
 since both $\phi$ and $\phi_k$ depend on $q$. In particular, for an anisotropic fermion band in the psedo gap phase, the spin gap can shift in some directions but not in others.  This implies that in the pseudo gap phase, the thermal population of  spin fluctuations should be dependent of the direction of $q$. Their contribution  to the scattering of the
  phonons or the charge degrees of freedom should be dependent on $q$. Moreover, for dispersion $\omega(q)=\alpha |q|$, a magnon with momentum 
 $q$ can decay into two magnons with momentum  $q_1+q_2=q$ in the same direction conserving both energy and momentum. Such a decay is no
 longer allowed if the lower frequencies are shifted more than the higher ones. 
 Therefore  anisotropic shifts would cause anisotropic effects on the thermal conductivity such as the ones observed experimentally.~\cite{ausloos99} 
 
 The appearance of spin gap has been related to the increase of thermal conductivity below Tc 
 in cuprates\cite{ausloos99} and picnitides.~\cite{may13} This increase is absent in conventional superconductors. 
 Within the context of this model, this is a direct consequence of the condition for an attractive interaction near the
 Fermi surface [see Eq. (\ref{eq:condsup})]. Low frequency and low momentum spin excitations appear to be involved in unconventional superconductivity, while in contrast in conventional superconductivity higher frequency phonon modes with momentum $q\simeq 2 k_F$ play a dominant role. Therefore, when unconventional superconductivity occurs (i) a spin gap appears, (ii)
 low energy and low momentum spin excitations disappear, and (iii) channels for low energy scattering are removed. This argument has been used to explain the increment of the thermal conductivity below T$_c$. In fact, the 
 behavior of the thermal conductivity has been used as a marker that distinguishes conventional and unconventional superconductors.~\cite{may13} 

 \section{The case of $p_\pi$}
 \label{sc:case}
 Since $p_\pi$ has shown the ability to generate a superconducting ground state in the presence of an antiferromagnetic
 background, the consideration of  the controversial question of whether $p_\pi$ is actually responsible for the superconductivity
 observed in cuprates cannot be avoided.  Is there any chance that $p_\pi$ deserves  credit for superconductivity? 
  
Emery is credited~\cite{emery87} for selecting the orbitals included in the three band Hubbard model which neglects $p_\pi$. 
 Emery's argument is based on the assumption that orbital hybridization is dominant. Emery supported his argument with electronic
 structure calculations based on the local density approximation (LDA) of density 
 functional theory (DFT).~\cite{mattheiss87,yu87}  While at the time, those calculations were ``state of the art", it is now well 
 known~\cite{pickett89,mcmahan90,filippetti03,singh10,rivero13} that they suffered from severe self-interaction errors, which are particularly common in transition metal 
oxides.~\cite{filippetti03,singh10,filippetti11} Self-interaction errors increase the energy of the occupied localized $d_{x^2-y^2}$ orbitals as compared to the delocalized $p$. The difficulty of correcting these errors and identifying the symmetry of the doping carriers was recognized in early work on the area.~\cite{mcmahan90}  The correction of self-interaction errors in a DFT context is still the subject of intense research.~\cite{dabo10,kraisler13}
   
If one takes the point of view that hybridization is less important than crystalline fields, a different ordering of levels appears. 
It was early realized by Adrian~\cite{adrian88} that if crystalline fields are dominant the orbital $p_\pi$ is favored. If one focuses on the O atom, 
the closest ions to each O$^{-2}$ in the plane are two Cu$^{+2}$.  
The $p_\sigma$ orbital of O$^{-2}$  aligned in the direction of 
the Cu-Cu distance must have lower crystalline field energy because of the cation Cu$^+2$ charge. In other words, $p_\sigma$ 
 takes the most advantage of the Coulomb attraction of the neighboring positive ions. 
The energies of the remaining two O$p$ orbitals are split by the tetragonal lattice. 
The $p$ orbital perpendicular to the plane ($p_z$) should lower its energy, relative to the one in the plane ($p_\pi$), 
because the CO$_2$ plane has a net $-2$ charge per unit cell. The orbital $p_z$ moves away from the negative charges in 
the direction of a positive plane. The remaining orbital in the plane $p_\pi$, in contrast, 
remains close to the negative ions and should have the highest energy. 
Thus, considering crystalline effects only, the highest energy orbital in the O atom is $p_\pi$, perpendicular to the Cu-Cu distance. Accordingly, since the lowest energy configuration for a hole is at the 
highest energy electron band,
one competing configuration for a hole to Emery's model is (i) in the plane and (ii) in an orbital perpendicular to the Cu-Cu distance.~\cite{adrian88}
Crystalline field effects 
decrease polynomially with the Cu-O distance while hopping terms decrease exponentially. Thus as the lattice expands crystalline 
effects must dominate. 

Since kinetic energy and crystalline effects compete, it is at least of academical interest to consider an hypothetical  case where crystalline effects on level splitting dominate.  Moreover,
experimental evidence shows\cite{chimaissem99} that superconductivity appears as the lattice is expanded. 
 Often the oxygen atoms
buckle, that is, move up and down in a direction perpendicular to the plane.  The buckling 
diminishes as the lattice is expanded. 
For zero bucking angle, the first neighbor  $p_\pi$ align and the hopping $t$ between first neighbors $p_\pi$  is maximized, which increases the Fermi velocity and the bandwidth of
the $p_\pi$ band.  

Despite of the influence of Emery's model in the literature, the role orbitals play is far from being
 settled. There is still debate on whether the O orbitals in the plane ($p_\sigma$ and $p_\pi$) are the only ones that play 
a role  both theoretically~\cite{sakakibara10,wang11} and experimentally~\cite{sakurai11}.  
On the experimental side, since the discovery of superconductivity,~\cite{bednorz86} there has been an intense 
debate on the symmetry of the hole band and the location of these holes~\cite{himpsel88,nucker89,eskes90} that continues 
until today.~\cite{sakurai11}   

Early experimental work on  superconducting cuprates~\cite{himpsel88,nucker89}  indicated that O in the plane is primarily the place where the holes go. In particular the orbitals in the plane were identified, but since the direction of the $p_\sigma$ and $p_\pi$ alternates, the experimental data is consistent with any one of them being occupied by holes.~\cite{nucker89} 
 However, early photoelectric data on CuO,~\cite{eskes90} suggested that the holes would occupy primarily the
$d_{x^2-y^2}$ orbital. While Cu in CuO has a local configuration similar to the cuprates, O has instead four Cu neighbors in a crystalline field of 
$T_d$ symmetry.~\cite{asbrink69}  Therefore, the environment of O in CuO is qualitatively very different to the one in superconducting cuprates. 
Despite these differences Ref. \onlinecite{eskes90} is still considered relevant by a few theorists today. While initially the occupation of the state with $d_{x^2-y^2}$ 
symmetry was  presumed to be responsible for superconductivity,~\cite{emery87} Compton experiments\cite{sakurai11} indicated recently that the occupation of
this orbital only occurs in the over doped regime and thus, might be responsible for the collapse of the superconducting state. 

All this controversy suggests that a new generation of experimental methods and new theories that can systematically overcome
self-interaction errors  are required. Ab-initio quantum Monte Carlo (QMC) methods are beginning  to be used in transition metal 
oxides~\cite{kolorenc08,kolorenc10} and metals~\cite{hood12}
 and offer some
hope to settle the controversy at least on theoretical grounds. While
these QMC methods cannot be used in the short term to decide if a material is a superconductor, they can be used to determine 
which model Hamiltonian describes the essential physics.   Some of those calculations are currently under preparation
in a couple of QMC groups.  

\section{Summary and discussion}
\label{sc:summary}
A model of a metallic band in an antiferromagnetic background has been studied analytically. The theory  is 
 related to the spin fermion models studied by Pines, Schrieffer and collaborators~\cite{monthoux92,schmalian99,schrieffer88} and related literature. But instead of 
 coupling the charge degrees of freedom with the spin fluctuations of the same band, different bands are coupled.
 This model corresponds to a limiting case in which two bands, one an antiferromagnetic insulator and the other non-magnetic
metal, are coupled only via the exchange interaction. 
  It has been found that 
under certain conditions this model could result in a superconducting ground state similar to the BCS ground state found
with phonons. 

The microscopic interaction studied in this work also originates additional terms. This additional terms contribute to superconductivity. They introduce a  shift on the magnon frequencies at Tc. The magnon frequency shift is analogous to the phonon renormalization of self energies.~\cite{zeyher90} In this model, in the long wave limit, a spin gap appears that prevents inelastic scattering below an energy threshold in the superconducting state. 

While the present model describes some key qualitative features of the superconducting cuprates, the matter of whether
this is an accurate representation  is subject  to debate. This model would describe hole-doped  cuprates if a perpendicular 
oxygen orbital is occupied by holes but not the one which is in general considered (the one that mediates the antiferromagnetic superexchange). This model could represent the electron doped superconductors if the electrons
remain in the reservoir region (e.g. in the Nd cations on the NdCuO$_4$ system).

However, this model may also represent  materials yet to be discovered. 
In general, in an arbitrary antiferromagnet, a different orbital might have higher energy than the orbital that mediates the supercharge antiferromagnetic interaction. In that case, a different band will be populated by holes or electrons after doping. 
Potentially, this mechanism could occur in the particular case of the cuprates under certain conditions when the crystalline field is dominant.

This model predicts a superconducting transition when the Fermi velocity in the metallic band is larger than the speed of
spin excitations for $q \rightarrow 0$. This must happen on the entire Fermi surface. Otherwise, for an anisotropic fermionic band, a pseudo gap phase  could appear if the condition for a gap is satisfied in parts of the surface but not in others. 
This relationship between spin wave velocity, Fermi velocity and superconductivity can, in principle, 
be tested experimentally and used to decide if a given superconducting material, e.g. the cuprate superconductors, are a member of the class represented by this model or not. 

This paper predicts that a material where a metallic nonmagnetic band and antiferromagnetism coexist can be a superconductor. This opens the search for additional antiferromagnetic materials where superconductivity could be found. For this class of materials, the layered structure does not play an important role, but symmetry and orbital alignment does. As long as the metallic band that appears after doping is placed at symmetrical position with respect to opposite spins, and the 
spin-excitation velocity is small, superconductivity could occur also in three dimensional antiferromagnetic structures. 

The critical temperature will be determined by the magnitude of the exchange coupling $J_2$ and the volume in 
reciprocal space where the coupling is significant. Superconductors could be found by band engineering. Unfortunately, current
electronic structure methods are not reliable in the scale of energies required to design the band structure of a 
superconducting materials $0.01$ eV. This suggests that theoretical research focused in overcoming self-interaction errors of DFT approximations or methods that go beyond DFT must be encouraged and stimulated.  

Within the context of this model both phonons and spin excitations can contribute to a BCS-like superconducting state. Therefore, 
they can in principle conspire to increase the critical temperature. In addition, as long as there are electron-phonon and electron-magnon scattering terms, both phonons and magnos become coupled with the superconducting state. Therefore the frequencies of phonons
can be modified\cite{hadjiev98} at the superconducting transition even in the case where an electronic mechanism is responsible for the 
superconducting state. 

\acknowledgments The author is highly grateful to E. Dagotto, A. Moreo, S. Okamoto, T. Maier and B. Sales 
for discussions and C. Balseiro and B. Alascio and C. Proetto for their
long appreciated and remembered classes on BCS theory, magnetism and solid state. This work was supported by the Materials Science and Engineering division of Basic Energy Sciences, Department of Energy. 

\appendix
\section{Expresion for the linear transformation operator $P_1$}
\label{ap:p1}
Excluding terms $h_1$ in $H_{s-f}$ with $[h_1,H_0]$ one  obtains:
\begin{align}
\label{eq:s1}
P_1= & \frac{1}{2N_k} \left\{ 
\gamma \left(\frac{{q_1}-{q_2}}{2}\right)\left[{U_v} ({q_1},{q_2})\left(\frac{c^\dagger_{\downarrow,{k_1}+{q_2}-{q_1}} c_{\downarrow,{k_1}} a_{\alpha,{q_2}} a_{\beta,{q_1}}}{\epsilon (\downarrow,{k_1})+\omega (\alpha ,{q_2})+\omega (\beta ,{q_1})-\epsilon (\downarrow,{k_1}-{q_1}+{q_2})}+  \right.\right. \right.  \\
 \nonumber   
 & \;\;\;\;\;\;\;\;\;\;\;\;\;\;\;\;\;\;\;\;\;\;\;\;\;\;\;\;\;\;\;\;\;\;\;\;\;\;\;\;\;\;\;\;\;\;\;\; + \frac{c^\dagger_{\uparrow,{k_1}+{q_2}-{q_1}} c_{\uparrow,{k_1}} a^\dagger_{\beta,{q_2}} a^\dagger_{\alpha,{q_1}}}{\epsilon (\uparrow,{k_1})-\omega (\beta ,{q_2})-\omega (\alpha ,{q_1})-\epsilon (\uparrow,{k_1}-{q_1}+{q_2})}+  \\ 
\nonumber   & \;\;\;\;\;\;\;\;\;\;\;\;\;\;\;\;\;\;\;\;\;\;\;\;\;\;\;\;\;\;\;\;\;\;\;\;\;\;\;\;\;\;\;\;\;\;\;\; - \frac{c^\dagger_{\downarrow,{k_1}+{q_2}-{q_1}} c_{\downarrow,{k_1}} a^\dagger_{\beta,{q_2}} a^\dagger_{\alpha,{q_1}}}{\epsilon (\downarrow,{k_1})-\omega (\beta ,{q_2})-\omega (\alpha ,{q_1})-\epsilon (\downarrow,{k_1}-{q_1}+{q_2})}+  
\\ \nonumber   & \;\;\;\;\;\;\;\;\;\;\;\;\;\;\;\;\;\;\;\;\;\;\;\;\;\;\;\;\;\;\;\;\;\;\;\;\;\;\;\;\;\;\;\;\;\;\;\; \left . - \frac{c^\dagger_{\uparrow,{k_1}+{q_2}-{q_1}} c_{\uparrow,{k_1}} a_{\alpha,{q_2}} a_{\beta,{q_1}}}{\epsilon (\uparrow,{k_1})+\omega (\alpha ,{q_2})+\omega (\beta ,{q_1})-\epsilon (\uparrow,{k_1}-{q_1}+{q_2})} \right)+ \\ 
& \nonumber \;\;\;\;\;\;\;\;\;\;\;\;\;\;\;\;\;\;\;\;\;\;\;\;\;\;\;\;\;+ {U_u}({q_1},{q_2}) \left(\frac{c^\dagger_{\uparrow,{k_1}+{q_1}-{q_2}} c_{\uparrow,{k_1}} a^\dagger_{\alpha,{q_2}} a_{\alpha,{q_1}}}{\epsilon (\uparrow,{k_1})+\omega (\alpha ,{q_1})-\omega (\alpha ,{q_2})-\epsilon (\uparrow,{k_1}+{q_1}-{q_2})}+ \right.  \\ 
\nonumber   & \;\;\;\;\;\;\;\;\;\;\;\;\;\;\;\;\;\;\;\;\;\;\;\;\;\;\;\;\;\;\;\;\;\;\;\;\;\;\;\;\;\;\;\;\;\;\;\; + \frac{c^\dagger_{\downarrow,{k_1}+{q_2}-{q_1}} c_{\downarrow,{k_1}} a^\dagger_{\beta,{q_2}} a_{\beta,{q_1}}}{\epsilon (\downarrow,{k_1})+\omega (\beta ,{q_1})-\omega (\beta ,{q_2})-\epsilon (\downarrow,{k_1}-{q_1}+{q_2})}+ 
 \\ \nonumber   & \;\;\;\;\;\;\;\;\;\;\;\;\;\;\;\;\;\;\;\;\;\;\;\;\;\;\;\;\;\;\;\;\;\;\;\;\;\;\;\;\;\;\;\;\;\;\;\; - \frac{c^\dagger_{\uparrow,{k_1}+{q_2}-{q_1}} c_{\uparrow,{k_1}} a^\dagger_{\beta,{q_2}} a_{\beta,{q_1}}}{\epsilon (\uparrow,{k_1})+\omega (\beta ,{q_1})-\omega (\beta ,{q_2})-\epsilon (\uparrow,{k_1}-{q_1}+{q_2})}+  \\ 
\nonumber   & \left.\left. \;\;\;\;\;\;\;\;\;\;\;\;\;\;\;\;\;\;\;\;\;\;\;\;\;\;\;\;\;\;\;\;\;\;\;\;\;\;\;\;\;\;\;\;\;\;\;\; - \frac{c^\dagger_{\downarrow,{k_1}+{q_1}-{q_2}} c_{\downarrow,{k_1}} a^\dagger_{\alpha,{q_2}} a_{\alpha,{q_1}}}{\epsilon (\downarrow,{k_1})+\omega (\alpha ,{q_1})-\omega (\alpha ,{q_2})-\epsilon (\downarrow,{k_1}+{q_1}-{q_2})}\right)\right] + \\ & \nonumber 
-\sqrt{2} N_k \sqrt{\frac{S}{N_k}} {U_p}({q_1}) \gamma \left(\frac{{q_1}}{2}\right)
\left(\frac{c^\dagger_{\uparrow,{k_1}-{q_1}} c_{\downarrow,{k_1}} a^\dagger_{\alpha,{q_1}}}{\epsilon (\downarrow,{k_1})-\omega (\alpha ,{q_1})-\epsilon (\uparrow,{k_1}-{q_1})}+ \right. \\ \nonumber   & \;\;\;\;\;\;\;\;\;\;\;\;\;\;\;\;\;\;\;\;\;\;\;\;\;\;\;\;\;\;\;\;\;\;\;\;\;\;\;\;\;\;\;\;+
\frac{c^\dagger_{\downarrow,{k_1}+{q_1}} c_{\uparrow,{k_1}} a_{\alpha,{q_1}}}{\epsilon (\uparrow,{k_1})+\omega (\alpha ,{q_1})-\epsilon (\downarrow,{k_1}+{q_1})}+  \\ 
 \nonumber      &\;\;\;\;\;\;\;\;\;\;\;\;\;\;\;\;\;\;\;\;\;\;\;\;\;\;\;\;\;\;\;\;\;\;\;\;\;\;\;\;\;\;\;\; + \frac{c^\dagger_{\downarrow,{k_1}+{q_1}} c_{\uparrow,{k_1}} a^\dagger_{\beta,{q_1}}}{\epsilon (\uparrow,{k_1})-\omega (\beta ,{q_1})-\epsilon (\downarrow,{k_1}+{q_1})}+  \\  \nonumber      
& \left.\left.  \;\;\;\;\;\;\;\;\;\;\;\;\;\;\;\;\;\;\;\;\;\;\;\;\;\;\;\;\;\;\;\;\;\;\;\;\;\;\;\;\;\;\;\;  +
\frac{c^\dagger_{\uparrow,{k_1}-{q_1}} c_{\downarrow,{k_1}} a_{\beta,{q_1}}}{\epsilon (\downarrow,{k_1})+\omega (\beta ,{q_1})-\epsilon (\uparrow,{k_1}-{q_1})}
\right)\right\}. 
\end{align}

\section{Expressions for $V_{a^\dagger, a^\dagger}$ and $V_{a, a}$ }
\label{ap:adad}
 \begin{align}
&V_{a^\dagger, a^\dagger} (k_1,k_2,q_1)= \frac{1}{8} J_2^2 \frac{1}{N_k^2} \gamma \left(\frac{{q_1}}{2}\right)^2
\left\{
U_u({q_2}-{q_1},{q_2})
\left[U_v({q_2}-{q_1},{q_2}) \right.\right. \\ 
\nonumber &
\left(\frac{1}{-\epsilon ({k_1}-{q_1})+\epsilon ({k_1})+\omega ({q_2}-{q_1})-\omega ({q_2})}-\frac{1}{-\epsilon ({k_2}+{q_1})+\epsilon ({k_2})-\omega ({q_2}-{q_1})-\omega ({q_2})}\right)+ 
\\  & + \nonumber U_v({q_2},{q_2}-{q_1})\\  & 
\nonumber \left. \left( \frac{1}{-\epsilon ({k_1}-{q_1})+\epsilon ({k_1})-\omega ({q_2}-{q_1})-\omega ({q_2})}-\frac{1}{-\epsilon ({k_2}+{q_1})+\epsilon ({k_2})+\omega ({q_2}-{q_1})-\omega ({q_2})}\right) \right]+ \\ 
 & + \nonumber U_u({q_1}+{q_2},{q_2}) \left[U_v({q_2},{q_1}+{q_2}) \right. \\  & 
 \nonumber \left( \frac{1}{-\epsilon ({k_2}+{q_1})+\epsilon ({k_2})-\omega ({q_1}+{q_2})-\omega ({q_2})}-\frac{1}{-\epsilon ({k_1}-{q_1})+\epsilon ({k_1})+\omega ({q_1}+{q_2})-\omega ({q_2})} \right)+ \\  & 
 + \nonumber U_v({q_1}+{q_2},{q_2})\\  & 
 \left.\left.
 \nonumber \left( \frac{1}{-\epsilon ({k_2}+{q_1})+\epsilon ({k_2})+\omega ({q_1}+{q_2})-\omega ({q_2})}-\frac{1}{-\epsilon ({k_1}-{q_1})+\epsilon ({k_1})-\omega ({q_1}+{q_2})-\omega ({q_2})}\right)\right] \right\}  
\end{align}
\begin{align}
&V_{a,a} (k_1,k_2,q_1)= 
\frac{1}{8} J_2^2 \frac{1}{N_k^2} \gamma \left(\frac{{q_1}}{2}\right)^2
\left\{ U_u({q_2},{q_2}-{q_1}) \left[ U_v({q_2},{q_2}-{q_1})  \right.\right. \\ & \nonumber 
\left(\frac{1}{-\epsilon ({k_2}+{q_1})+\epsilon ({k_2})-\omega ({q_2}-{q_1})+\omega ({q_2})}-\frac{1}{-\epsilon ({k_1}-{q_1})+\epsilon ({k_1})+\omega ({q_2}-{q_1})+\omega ({q_2})}\right)+ 
\\  & + \nonumber U_v({q_2}-{q_1},{q_2}) \\  &  
\nonumber \left.
\left(\frac{1}{-\epsilon ({k_2}+{q_1})+\epsilon ({k_2})+\omega ({q_2}-{q_1})+\omega ({q_2})}-\frac{1}{-\epsilon ({k_1}-{q_1})+\epsilon ({k_1})-\omega ({q_2}-{q_1})+\omega ({q_2})}\right)\right]+ 
\\  & + \nonumber U_u({q_2},{q_1}+{q_2})\left[U_v({q_1}+{q_2},{q_2}) \right. \\  & 
 \nonumber \left(\frac{1}{-\epsilon ({k_1}-{q_1})+\epsilon ({k_1})+\omega ({q_1}+{q_2})+\omega ({q_2})}-\frac{1}{-\epsilon ({k_2}+{q_1})+\epsilon ({k_2})-\omega ({q_1}+{q_2})+\omega ({q_2})}\right)+ \\  & + 
 \nonumber U_v({q_2},{q_1}+{q_2}) \\  & 
 \left.\left.
  \nonumber \left(\frac{1}{-\epsilon ({k_1}-{q_1})+\epsilon ({k_1})-\omega ({q_1}+{q_2})+\omega ({q_2})}-\frac{1}{-\epsilon ({k_2}+{q_1})+\epsilon ({k_2})+\omega ({q_1}+{q_2})+\omega ({q_2})}\right) \right] \right\}.
  \end{align}

\end{document}